\title{\textbf{Surface-mediated reduction of ion-irradiation-induced damage in tungsten revealed by advanced ion channeling analysis}}

\newcommand{\abstractText}{\noindent Tungsten is a leading candidate material for plasma-facing components in future fusion reactors. In this work, we integrate advanced ion channeling analysis with large-scale molecular dynamics simulations to uncover a pronounced surface-mediated reduction of radiation damage in single-crystal tungsten at elevated temperatures. We demonstrate how our unique analysis method can clearly resolve a dislocation-free zone near the surface and a transition region with suppressed defect density before reaching the bulk value. A possible explanation for the strong surface effect at elevated temperatures can be obtained by considering the coherent drift motion of dislocation loops toward the surface. \\

\noindent \textbf{Keywords}: Radiation effects, surface effects, ion channeling, dislocations, tungsten
}

\documentclass[12pt, a4paper, twocolumn]{article}

\usepackage[utf8]{inputenc} 
\usepackage{authblk} 

\author[1,*]{Xin Jin}
\author[2]{Fredric Granberg}
\author[2]{Kai Nordlund}
\author[3]{Sabina Markelj}
\author[3]{Esther Punzón-Quijorna}
\author[2]{Flyura Djurabekova}

\affil[1]{\normalsize\textit{Shenzhen Key Laboratory of Nuclear and Radiation Safety, Shenzhen University, Shenzhen 518060, China}}
\affil[2]{\normalsize\textit{Department of Physics, University of Helsinki, P. O. Box 43, Helsinki FIN-00014, Finland}}
\affil[3]{\normalsize\textit{Jožef Stefan Institute (JSI), Ljubljana, Slovenia}}

\affil[*]{Corresponding author: xin.tlg.jin@outlook.com}

\date{}
\usepackage[pdfborder={0 0 0}]{hyperref} 
\usepackage{graphicx} 
\usepackage{float} 
\usepackage{subcaption} 
\usepackage[labelfont=bf]{caption} 
\usepackage{indentfirst} 
\usepackage[top=1.2in, bottom=1.25in, left=0.9in, right=0.9in]{geometry}
\usepackage{amsmath} 
\usepackage{amssymb}
\usepackage{bm} 
\usepackage[T1]{fontenc} 
\usepackage{mathtools}
\usepackage{siunitx} 
\usepackage{gensymb} 
\usepackage{tabularx} 
\usepackage{booktabs} 
\usepackage{mhchem} 
\usepackage[sort&compress, numbers]{natbib} 
\usepackage{abstract}
\usepackage{miller}
\usepackage[normalem]{ulem} 

\DeclareUnicodeCharacter{2010}{-}
\DeclareUnicodeCharacter{01E3F}{\'{e}}
\DeclareUnicodeCharacter{03B1}{$\alpha$} 
\DeclareUnicodeCharacter{03B3}{$\gamma$} 
\DeclareUnicodeCharacter{0141}{\L{}} 

\begin{document}

\twocolumn[
  \begin{@twocolumnfalse}
    \maketitle
    \begin{abstract}
      \abstractText
      \newline
      \newline
    \end{abstract}
  \end{@twocolumnfalse}
]

\raggedbottom
%

The extreme operational conditions of future fusion reactors demand plasma-facing materials that can withstand intense heat fluxes and severe radiation damage \cite{Bolt2002, Schmid2024}. Currently, one of the leading candidates for plasma-facing materials is tungsten due to its highest melting point of all metals, low sputtering yield, and low tritium retention \cite{Philipps2011, Rieth2013}. Under irradiation, high-energy particles can displace tungsten atoms, causing damage that alters the properties of the material. Hence, extensive research has been conducted on radiation effects in tungsten, with particular emphasis on defect production and evolution \cite{Yi2016, Atwani2018, Mason2021, Wang2023a, Markelj2024}. 

To understand these irradiation-induced phenomena, a thorough characterization of irradiation-induced defects is essential, but remains non-trivial. Recently, a comprehensive experimental study of self-ion irradiation of single-crystal tungsten showed that, at a damage level of 0.02 dpa, increasing the irradiation temperature from \mbox{290 K} to \mbox{800 K} significantly lowers the signals of Rutherford backscattering spectrometry in channeling mode (RBS/c) \cite{Zavasnik2025}. RBS/c is a special ion beam analysis technique that is very sensitive to several types of crystal defects \cite{Gemmell1974, Feldman1982, Zhang2015c}. Therefore, the pronounced decrease in the RBS/c signal indicates a substantial reduction in radiation damage at \mbox{800 K}. Understanding the mechanism underlying this reduction in radiation damage is essential, since future fusion reactors will operate at elevated temperatures \cite{Bolt2002}. Nevertheless, it is difficult to unambiguously determine the nature of the radiation defects formed in tungsten solely on the basis of experiments without advanced simulations.

In this Letter, we reveal the mechanisms behind the reduction of radiation damage in tungsten, as observed in the RBS/c experiments \cite{Zavasnik2025}, using state-of-the-art simulation methods. We characterize the depth distribution of radiation defects, particularly dislocation loops, and demonstrate how a surface can generate a pronounced effect on the dislocation distribution at elevated temperatures. We demonstrate that this strong surface effect can be explained by temperature enhanced drift motion of dislocation loops, offering new insights into the relationship between defect mobility and resistance of materials to radiation damage. 

A series of RBS/c experiments \cite{Zavasnik2025} were performed on the \hkl<111>-oriented single-crystal tungsten samples irradiated with 10.8 MeV tungsten ions at different conditions (for details of the experiments, see Section 1 of Supplementary Materials). In this work, we chose to interpret the specific RBS/c spectra generated from the tungsten samples irradiated to \mbox{0.02 dpa} at \mbox{290 K} and \mbox{800 K}, in which the energy of probing ions, i.e., $^4$He ions, was \mbox{3 MeV}. For this purpose, we performed RBS/c simulations using the RBSADEC code \cite{Zhang2016, Jin2020, RBSADEC_ND}, which can generate signals directly comparable to experiments. A critical prerequisite in the simulation is to supply the RBSADEC code with simulated targets accurately representing realistic atomic structures. To this regard, molecular dynamics (MD) simulations were performed to mimic radiation effects in the experiments. Radiation damage was created by overlapping collision cascades initiated by a \mbox{10 keV} primary knock-on atom, similarly to Ref. \citenum{Granberg2021}. The time interval between cascades was 30 ps. Periodic boundary conditions were applied to all the directions. A newly developed potential for describing the behavior of tungsten under irradiation \cite{Mason2023} was used in all the steps of MD simulations. The threshold displacement energy of tungsten atom was set to 90 eV \cite{Banisalman2017}. The dpa production rate was calculated to be $2.45 \times 10^6$ dpa/s. After the generation of radiation damage, dozens of MD cells were connected along the \hkl<111> direction and were relaxed in an additional MD simulation, which finally produced a single super-cell, whose size along the \hkl<111> direction is 1.4 $\mu$m. 

Our previous study \cite{Markelj2024} demonstrated that simulated RBS/c spectra from this type of super-cell can well reproduce experimental signals at 0.02 dpa and at room temperature (RT). In this study, we further improved the connection process leading to that all MD cells can be perfectly connected. The overall connection procedure aims to prevent the generation of artificial defects at the interfaces between neighboring MD cells. The MD cells, which retain periodic boundary conditions in all directions, were carefully shifted to ensure defect-free interfaces and proper alignment of atomic rows, followed by a final relaxation of the entire system for 60 ps, using the LAMMPS code \cite{Plimpton1995, Thompson2022}.

When connecting the MD cells, each MD cell can have different numbers of collision cascades and hence different levels of radiation damage. The depth profile of cascade number can be determined in such a way that the depth profile of damage dose in dpa corresponds to that calculated by the SRIM software, which is based on the binary collision approximation (BCA) \cite{Ziegler_ND, Ziegler2015}. Hereafter, we refer to this methodology as the BCA-MD approach for the super-cell generation. An alternative approach involves manually adjusting the depth profiles of MD cells to directly fit the experimental spectra. Hereafter, we refer to this approach as the Fit-MD approach, in which the arrangement of MD cells with different number of collision cascades was iteratively adjusted until the simulated RBS/C spectra reproduce the experimental signals over the entire depth range. (For details of RBS/c and MD simulations, see Section 1 and Section 2 of Supplementary Materials).

Fig. \ref{fig:rbs_1}(a) shows the experimental and simulated RBS/c signals obtained by using \mbox{3 MeV} He ion beams on \hkl<111>-oriented tungsten samples, which are represented by the dots and lines, respectively. The simulations are based on the BCA-MD approach. The simulated result at RT agrees very well with the experimental one, with only a small yield overestimation in the high energy region (corresponding to the surface region) as shown in the inset. Good agreement has also been observed between MD simulations and experiments in the measurement of deuterium retention \cite{Mason2021}, thermal diffusivity of irradiated tungsten \cite{Reza2020} and Positron Annihilation Spectroscopy studies \cite{hu2025}. Our RBS/c results further demonstrate that this agreement is consistent throughout almost the entire depth profile.

The yield of the experimental RBS/c spectrum of the sample irradiated at \mbox{800 K} is almost half that of the sample irradiated at RT, as shown in Fig. \ref{fig:rbs_1}(a). This substantial decrease in RBS/c signals provides a strong indication that irradiation at elevated temperatures results in a reduction in radiation damage. Furthermore, at this elevated temperature, we observe a significant discrepancy between the simulated and experimental spectra. In fact, the simulated spectra at \mbox{800 K} remain almost the same as those at RT. To understand the difference in the radiation effects in tungsten irradiated at different temperatures, we quantitatively analyzed the atomistic structures obtained by means of BCA-MD with predominant radiation defects, namely 1/2\hkl<111> dislocation loops, which are commonly reported to form in tungsten under ion irradiation \cite{Ferroni2015, Durrschnabel2021, Markelj2024}. 

Fig. \ref{fig:rbs_1}(b) shows that the dislocation density at RT practically remains constant (\mbox{$\sim 0.95 \times 10^{16}$ m$^{-2}$}) throughout the damaged layer up to \mbox{860 nm} beneath the surface, then gradually decreases to zero by \mbox{1.3 $\mu$m}. (The depth resolution in this study is determined by the size of individual MD cells, i.e., $\sim$ 20 nm). For the density distribution at \mbox{800 K}, unlike the significant decrease of RBS/c signals, the general trend at \mbox{800 K} is similar to that at RT, except for the slightly lower dislocation density within the damaged region  ($\sim 0.80 \times 10^{16}$ m$^{-2}$). 
Hence, the discrepancy between the simulated and experimental RBS/c spectra measured at RT and at \mbox{800 K} is likely to stem from a factor not considered in the BCA-MD approach, for example, significant defect mobility at long time scales, inaccessible with conventional MD. In order to reproduce the experimental RBS/C signals at 800 K, it is necessary to manually adjust the depth distribution of MD cells containing different densities of dislocation loops until the simulated RBS/C spectra agree with the experimental signals. In addition, TEM observations \cite{Wang2024, Zavasnik2025} indicate that the nature of the dislocation loops does not undergo significant changes from RT to 800 K. Therefore, it is reasonable to use the present MD cells, which predominantly contain 1/2\hkl<111> dislocation loops, to fit the experimental results at 800 K. This is the main idea of the Fit-MD approach, which will be described in the following.

\begin{figure}[!h]
     \centering
     \begin{subfigure}[b]{0.45\textwidth}
         \centering
         \includegraphics[width=\textwidth]{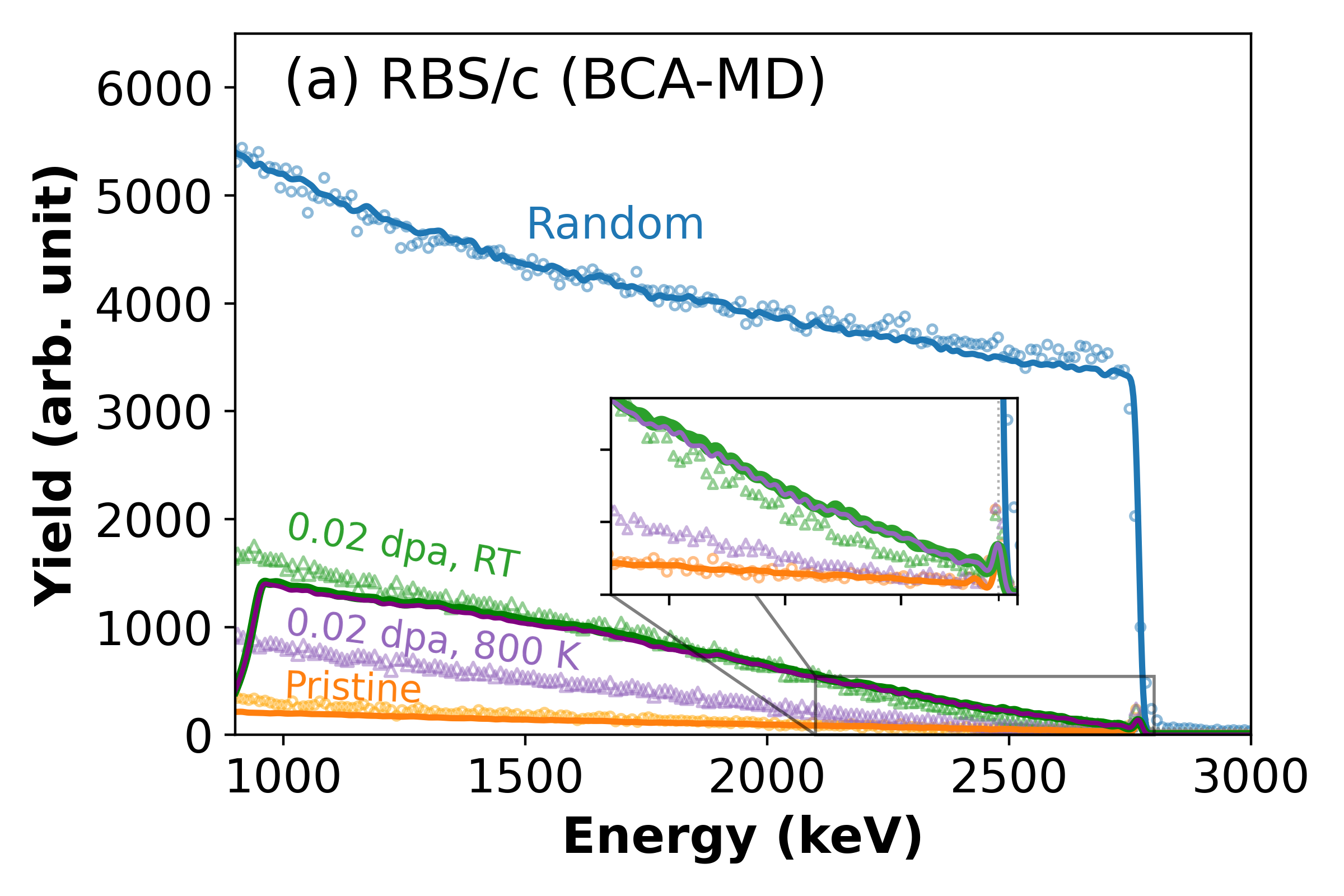}
     \end{subfigure}
     \begin{subfigure}[b]{0.45\textwidth}
         \centering
         \includegraphics[width=\textwidth]{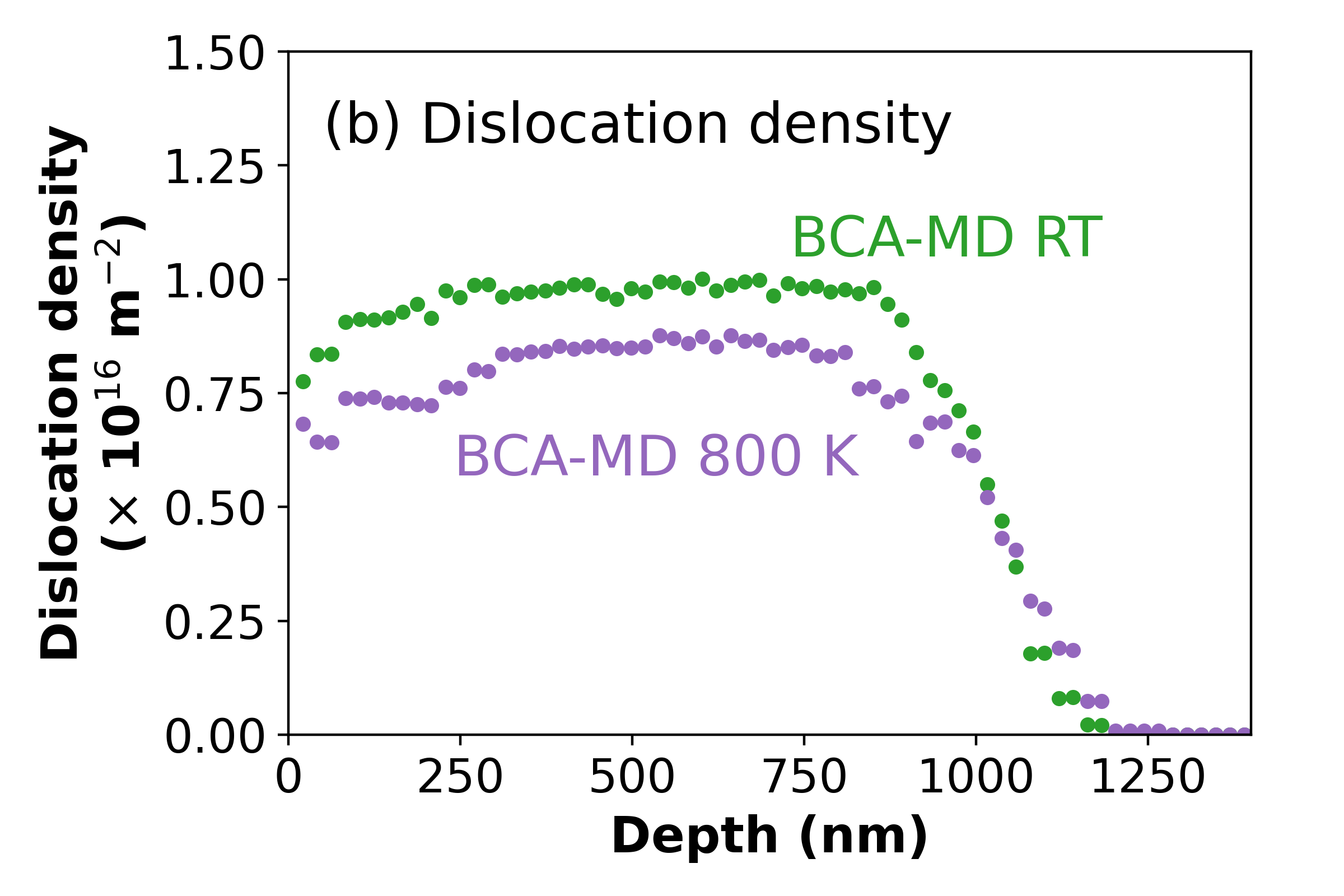}
     \end{subfigure}
     \caption{Analysis of RBS/c signals and radiation damage in tungsten using the BCA-MD approach: (a) Experimental (symbols) and simulated (lines) RBS/c spectra of \mbox{3 MeV} He ions on \hkl<111>-oriented tungsten samples at RT and \mbox{800 K} (maximum damage dose: \mbox{0.02 dpa}),  and (b) depth distribution of dislocation density in the simulated targets.}
     \label{fig:rbs_1}
\end{figure}

Using the Fit-MD approach, the simulated signals can perfectly reproduce the experimental ones at both temperatures, as shown in Fig. \ref{fig:rbs_fit}(a). In addition, we also fitted the pristine spectrum; for more details, see Section 2 of Supplementary Materials. The depth profiles of the dislocation density are presented in Fig. \ref{fig:rbs_fit}(b). In these results, we observe that the maximum of the dislocation density in the damaged samples, see Fig. \ref{fig:rbs_fit}(b), remains approximately the same as in the BCA-MD simulation cells shown in Fig. \ref{fig:rbs_1}(b). However, the variations in depth distributions of the dislocation density are much stronger in the Fit-MD approach. 

\begin{figure}[!h]
     \centering
     \begin{subfigure}[b]{0.45\textwidth}
         \centering
         \includegraphics[width=\textwidth]{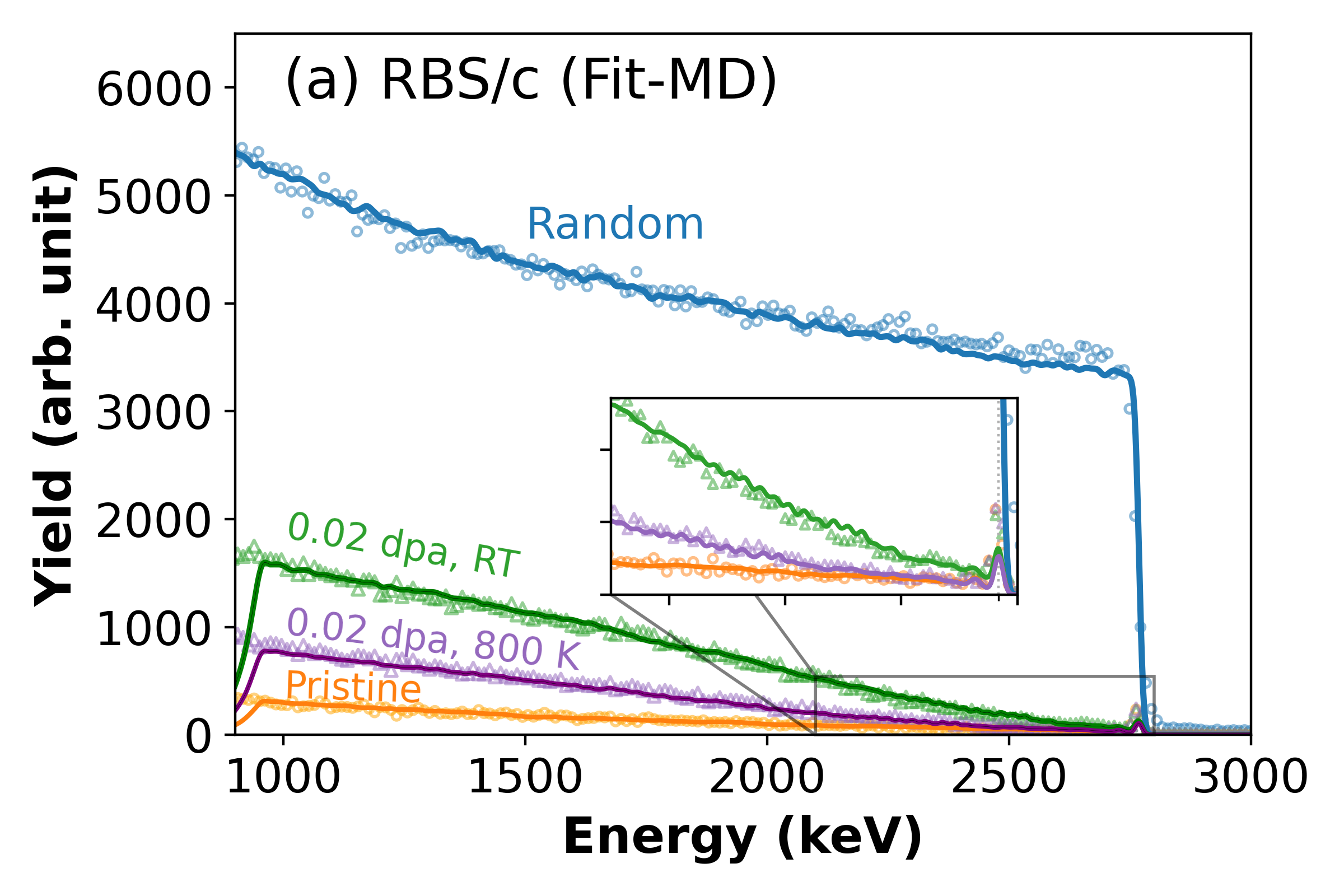}
     \end{subfigure}
     \begin{subfigure}[b]{0.45\textwidth}
         \centering
         \includegraphics[width=\textwidth]{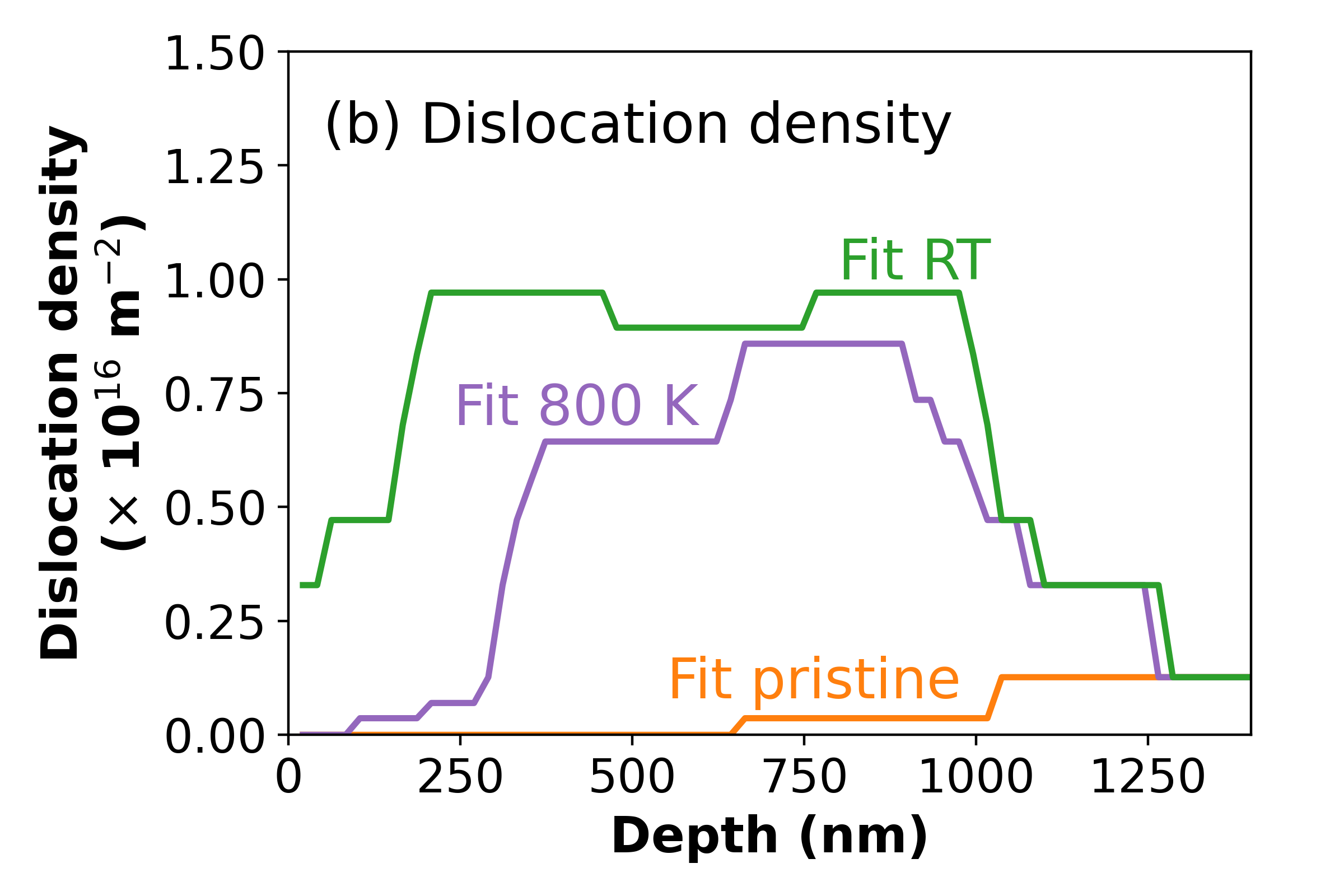}
     \end{subfigure}
     \caption{Analysis of RBS/c signals and radiation damage in tungsten using the Fit-MD approach: (a) Experimental (symbols) and simulated (lines) RBS/c spectra of \mbox{3 MeV} He ions on \hkl<111>-oriented tungsten samples at RT and \mbox{800 K} (maximum damage dose: \mbox{0.02 dpa}), and (b) depth distribution of dislocation density in the simulated targets.}
     \label{fig:rbs_fit}
\end{figure}

\begin{figure}[!h]
  \centering
  \includegraphics[width=0.85\linewidth]{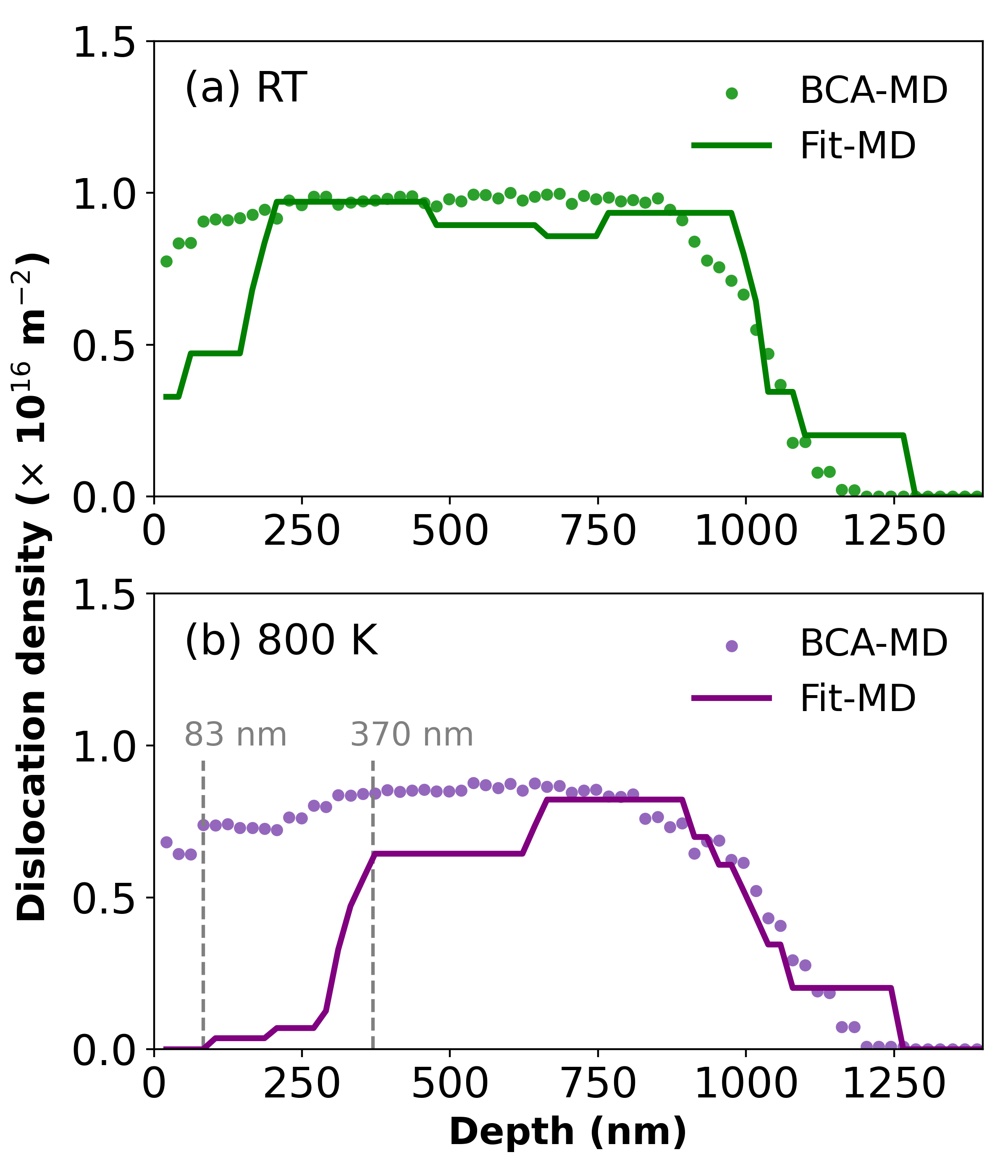}
  \caption{Comparison of dislocation densities obtained from the BCA-MD and Fit-MD approaches at (a) RT and (b) \mbox{800 K}.}
  \label{Fig:dist_compare}
\end{figure}

Figs. \ref{Fig:dist_compare}(a) and \ref{Fig:dist_compare}(b) compare the depth distribution of the dislocation density, which we obtained from the two approaches at RT and \mbox{800 K}, respectively. The Fit-MD results presented in Fig. \ref{Fig:dist_compare} are obtained by subtracting the density in the pristine sample from the damaged ones given in Fig. \ref{fig:rbs_fit}(b), thus excluding the influence of the signal from the pristine samples. At RT, both approaches yielded similar results with extended plateaus in the central regions of the depth profile. However, the Fit-MD results indicate that the defect density must decrease near the surface (< 200 nm). Indeed, the formation of a dislocation-denuded zone of 10 nm was observed in the TEM studies at RT \cite{Zavasnik2025}. This is consistent with calculations based on elasticity theory \cite{Jager1975} showing that when the distance between dislocation loops and the free surface is only a few times larger than the loop size, the surface is effective in attracting and annihilating dislocation loops.

Fig. \ref{Fig:dist_compare} (b) shows that, at \mbox{800 K}, there are significant differences between the two approaches. Although the result of the BCA-MD approach still shows a nearly constant dislocation density that extends from the surface to the bulk region ($\sim$ 900 nm), the result of the Fit-MD approach indicates a pronounced surface effect. We observe a dislocation-free zone with a thickness of 83 nm, which is much larger than the size of the dislocation loops. Conventionally, surface effects are mainly characterized by the width of such a defect-free zone. However, since the defect-free zone must be followed by a region where the defect density progressively increases from zero to the bulk value, the reduction in the defect density in this transition region should also be attributed to the surface effect \cite{Vigeholm1965, LiYong2023}. In our case, this transition region covers an extended depth, ranging from 83 nm to at least 370 nm, in which the dislocation density is substantially lower than that in the bulk region, where the defect distribution is nearly constant. A key challenge in characterization of the transition region lies in determination of the onset of the bulk region. By comparing the defect distribution in the two approaches, we are able to delineate the clear boundary between the transition and bulk regions ($\sim$ 370 nm), thus highlighting the capability of our RBS/c analysis. In the previous TEM study \cite{Zavasnik2025}, the depth distribution of the total dislocation density (including both lines and loops) did not clearly reveal such a strong surface influence. Instead, TEM results show that a defect-depleted zone decreases with increasing temperature (from $\sim$10 nm at 300 K to $\sim$6 nm at 800 K), which appears inconsistent with the expected enhancement of defect migration toward the surface at elevated temperatures. One possible reason is that TEM sample preparation using focused ion beam techniques could introduce artificial defects, affecting the interpretation of the surface effect. However, if we focus on the contribution given solely by dislocation loops as observed from TEM, we can find that their distribution is also significantly affected by the surface at \mbox{800 K}, showing a similar trend as observed in our RBS/c results.

\begin{figure}[!h]
  \centering
  \includegraphics[width=0.95\linewidth]{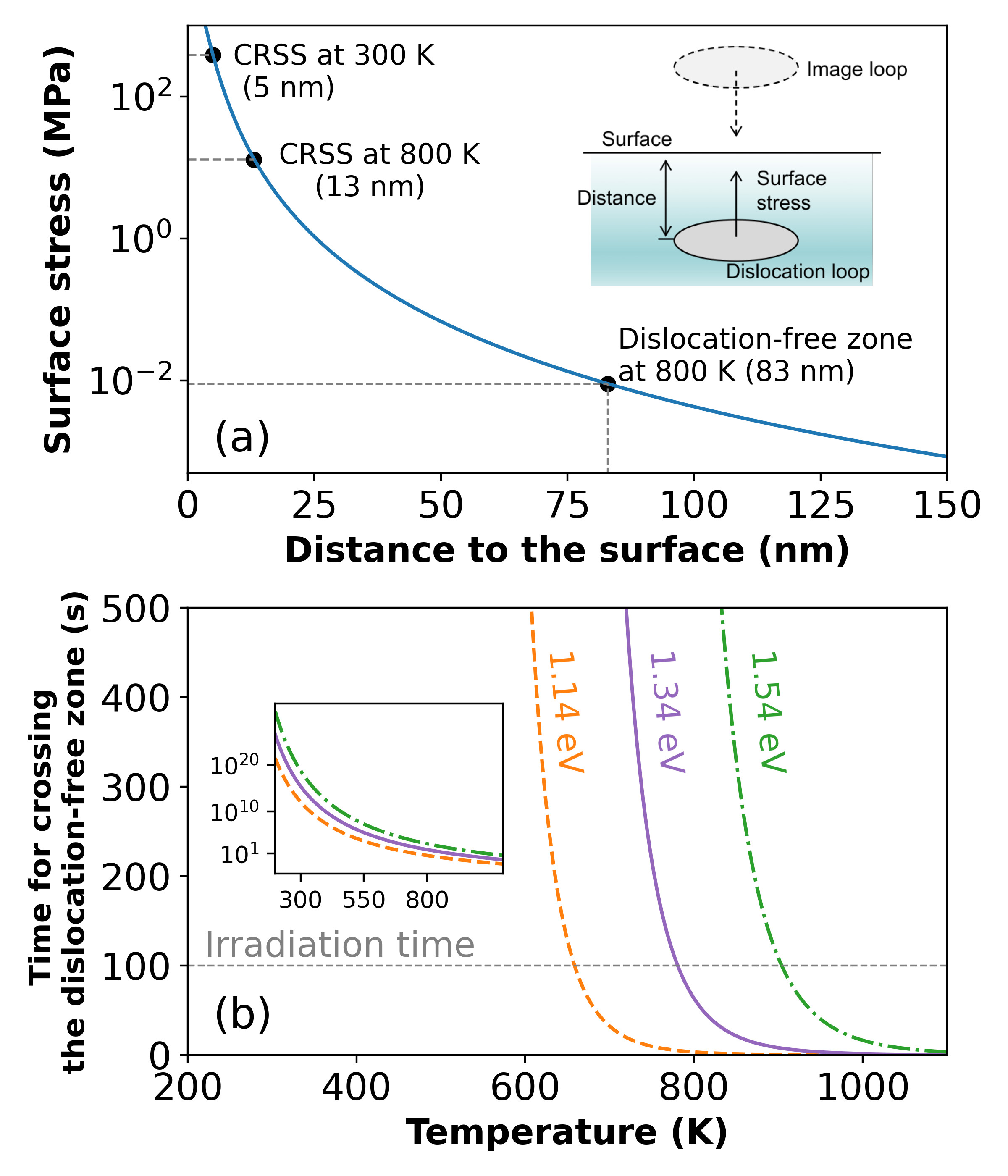}
  \caption{Evaluation of surface effects on 1/2\hkl<111> dislocation loops in tungsten: (a) Surface stress due to the vicinity of the open surface on the dislocation loop, and (b) time required for a dislocation loop to cross the dislocation-free zone (83 nm) as a function of temperature with different effective migration energies. The inset shows the corresponding time on a logarithmic scale.}
  \label{Fig:surf_effect}
\end{figure}

By extending to nearly a third of the damaged region (370 nm) that was produced in tungsten by the irradiation of 10.8 MeV self-ions, the surface effect appears to play a crucial role in determination of radiation defects in tungsten at \mbox{800 K}. Even the thickness of the dislocation-free zone (83 nm) is not a negligible value. This dislocation-free zone is much larger than the denuded zone observed by TEM at RT (10 nm) \cite{Zavasnik2025}, and is comparable or even greater than the grain size (35 - 100 nm) in a nanocrystalline tungsten samples, which show good irradiation tolerance \cite{Atwani2018, Atwani2019a}. It should be noted that, at elevated temperatures, where both interstitial- and vacancy-type defects become mobile \cite{Thompson1960, Eyre1973, Debelle2008}, we do not exclude the scenarios in which defect annihilation can also occur via interstitial-vacancy recombination. However, the surface effect should be dominant regarding the decrease in dislocation densities in the near-surface region.

To better understand the mechanisms underlying this pronounced surface effect, in particular, the formation of a dislocation-free zone, we conducted calculations to quantify how the surface affects dislocation loops. Fig. \ref{Fig:surf_effect}(a) presents the stress on a 1/2\hkl<111> circular dislocation loop in tungsten as a function of the distance between the loop and the open surface. The loop of 5 nm in diameter is parallel to the surface and its Burgers vector is parallel to the surface normal. The stress is calculated based on Baštecká's equation \cite{Bastecka1964, Fikar2015}. Conventionally, a defect-free zone is determined by comparing the stress applied to the loop with a critical slip stress \cite{Bastecka1964}, such as the critical resolved shear stress (CRSS) \cite{Jager1975a}. By comparing the stress exerted at the loop in Fig. \ref{Fig:surf_effect}(a) with CRSS of tungsten obtained from experiments \cite{Brunner2000}, we estimate the dislocation-free zone to be 5 and 13 nm at RT and \mbox{800 K}, respectively. Although the former is close to the TEM observation (10 nm), the latter is much lower than that obtained from our RBS/c analysis (83 nm). In fact, when the distance between the loop and the open surface is 83 nm, the stress on the loop ($\sim$ 0.01 MPa) is very small compared to the CRSS at \mbox{800 K} (\mbox{13 MPa}). Most likely, this discrepancy arises from the fact that the CRSS in BCC metals is mainly determined by screw dislocations \cite{Ohsawa2005, Clouet2021}. However, the irradiation-induced dislocation loops are mainly of edge character, and studies indicate that the mobility of edge dislocations is higher than that of screw dislocations \cite{Solomon1971, Vitek1973}.

Hence, instead of relying on a single critical stress to evaluate the dislocation-free zone, we propose to focus on the dynamic behavior of dislocation loops under the influence of external stress, i.e., drift motion of dislocation loops \cite{Hirth1982, Mehrer2007}. Under the force induced by the surface effect, $F$, the drift velocity of dislocation loops, $v_\text{D}$, can be calculated as follows:  
\begin{equation}
v_\text{D} = F \cdot u (T, b, d, \tau, E_\text{m}) \label{Eq:drift_v}
\end{equation}
where $u$ is the loop mobility, which depends on the temperature, $T$, the Burgers vector of the loop, $b$, the loop diameter, $d$, the external stress, $\tau = F/\pi d b$, and the effective migration energy, $E_\text{m}$. Note that the drift velocity $v_\text{D}$ increases as the loop approaches the open surface, due to the higher surface force. The time required for a loop to cross a distance, $x_s$, before reaching the surface can be calculated by $\int_0^{x_s} dx/v_D(x)$. Using $E_\text{m} = 1.34$ eV (from dislocation annealing experiments \cite{Ferroni2015}), we can calculate that a loop takes 63.6 s to cross the dislocation-free zone (\mbox{83 nm}) at \mbox{800 K}. For calculation details, see Section 4 of Supplementary Materials. In ion-irradiation experiments, reaching a damage dose of 0.02 dpa typically requires an irradiation time on the order of 100 s \cite{Markelj2024}. The similarity between these timescales supports the idea that the drift motion of dislocation loops at elevated temperatures could provide a plausible explanation for the significant surface effect. Fig. \ref{Fig:surf_effect}(b) demonstrates the rapid decrease of the time needed for a dislocation loop to cross the dislocation-free zone (83 nm) with a higher irradiation temperature. While this time is below 100 s at 800 K, it increases to $\sim 10^{15}$ s at 300 K. Hence, the temperature strongly affects the drift behavior of dislocation loops. 

In conclusion, we investigated the reduction of experimental RBS/c signals in single-crystal tungsten irradiated at elevated temperatures by combining advanced RBS/c and MD simulation methods. Our analysis clearly shows that raising the irradiation temperature from room temperature to \mbox{800 K} significantly lowers the near-surface dislocation density. 
This pronounced defect reduction can be attributed to a thermally enhanced surface effect, driven by the higher mobility of dislocation loops at elevated temperatures, which promotes the drift motion of loops towards the surface.

\section*{CRediT authorship contribution statement}

\textbf{Xin Jin}: Conceptualization, Data curation, Formal analysis, Investigation, Methodology, Software, Validation, Visualization, Writing – original draft. \textbf{Fredric Granberg}: Conceptualization, Formal analysis, Methodology, Software, Writing – original draft. \textbf{Kai Nordlund}: Conceptualization, Formal analysis, Funding acquisition, Supervision, Writing – original draft. \textbf{Sabina Markelj}: Conceptualization, Formal analysis, Funding acquisition, Methodology, Project administration, Writing – original draft. \textbf{Esther Punzón-Quijorna}: Conceptualization, Formal analysis, Methodology, Writing – original draft. \textbf{Flyura Djurabekova}: Conceptualization, Formal analysis, Funding acquisition, Project administration, Supervision, Writing – original draft.

\section*{Disclosure statement}

The authors declare that they have no known competing financial
interests or personal relationships that could have appeared to influence
the work reported in this paper.

\section*{Acknowledgements}

The authors wish to acknowledge CSC – IT Center for Science, Finland, for computational resources, and the Finnish Computing Competence Infrastructure (FCCI) for supporting this project with computational and data storage resources. This work was partially carried out within the framework of the EUROfusion Consortium, funded by the European Union via the Euratom Research and Training Programme (Grant Agreement No 101052200 — EUROfusion). Views and opinions expressed are however those of the author(s) only and do not necessarily reflect those of the European Union or the European Commission. Neither the European Union nor the European Commission can be held responsible for them. Xin Jin thanks to Shenzhen Science and Technology Program (ZDSYS20230626091501002) for financial support. This work is supported by the Intelligent Computing Center of Shenzhen University.

\bibliographystyle{naturemag}
\bibliography{./References/MyLibrary.bib,./References/Non_downloaded.bib}

\end{document}


\maketitle

\tableofcontents

\clearpage

\section{Experiments and simulations of Rutherford backscattering spectrometry in channeling mode}

Very recently, a series of RBS/c experiments had been performed on the \hkl<111>-oriented single-crystal tungsten samples irradiated with 10.8 MeV tungsten ions to 0.02 dpa and 0.2 dpa at different conditions, including multiple irradiation temperatures and multiple energies of probing ions \cite{Zavasnik2025}. In this study, we chose to interpret the RBS/c experiments generated from the tungsten samples irradiated to 0.02 dpa at 290 K and 800 K. The energy of probing ions, i.e., $^4$He ions, was 3 MeV. The probing direction is along the \hkl<111> direction. A detector located at 170\degree was used to detect backscattered He ions. The damage dose (0.02 dpa) was obtained based on SRIM calculations \cite{Ziegler_ND} using the "Quick calculation of damage" option. (For more details of the ion irradiation and RBS/c experiments, see Ref. \citenum{Zavasnik2025}.) The interpretation of spectra from samples irradiated to the higher dose of 0.2 dpa presents a greater technical challenge and will be addressed in future studies. 

\bigbreak

RBS/c simulations were performed to interpret the experimental signals using the RBSADEC code \cite{Zhang2016, Jin2020, RBSADEC_ND}. To better reproduce experimental signals, 3.017 MeV He ions were used to probe simulated tungsten targets along the \hkl<111> direction (Details of simulated targets will be presented in the next section). A detector with an energy resolution of 16.5 keV was located at 170\degree. Note that the RBS/c experiments were performed at room temperature (RT). Hence the temperature of simulated targets was set to 290 K. The Debye temperature of the simulated targets was set to 377 K \cite{Walford1969}, resulting in a one-dimensional thermal vibration magnitude of 4 pm for tungsten atoms. To reproduce the pristine spectra from experiments, the angular distribution of incoming ions was set to follow a Rayleigh distribution \cite{Jin2021b} with a characteristic angle of 0.32 \degree. In simulating the slowing down process of backscattered ions, a TRIM-exponential model was used to evaluate impact parameters in binary collisions. The distance of free flight path between two consecutive collisions was set to a relatively small value (10 nm) to prevent formation of artificial peaks in RBS/c spectra. To ensure that the detector can efficiently collect backscattered ions which have experienced significant multiple scattering events, the spread angle of backscattered ions was set to 60\degree. The electronic stopping power of He ions in tungsten was obtained from SRIM calculations.

\section{Molecular dynamics simulations}

\begin{figure}[h!]
  \centering
  \includegraphics[width=0.6\linewidth]{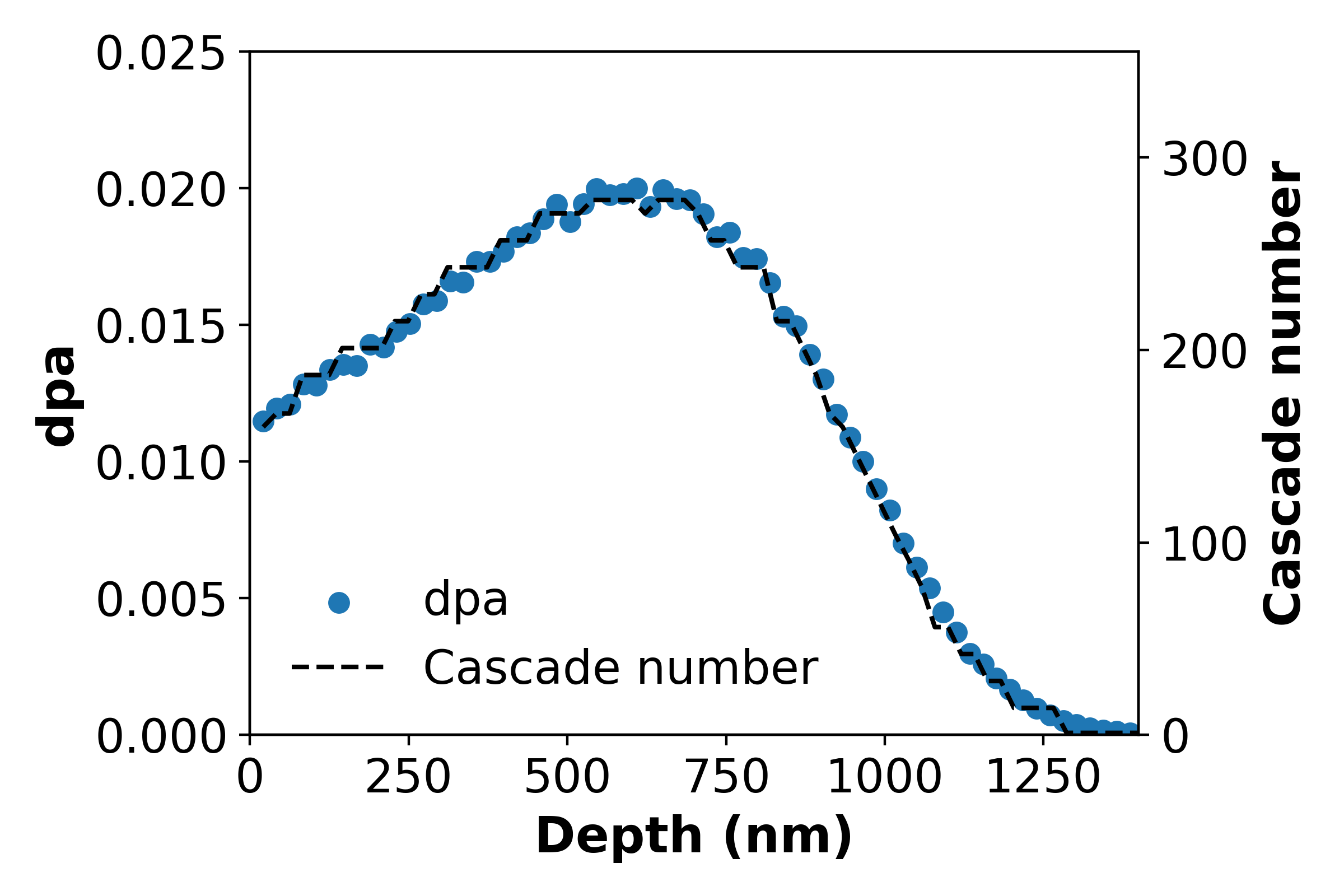}
  \caption{Depth distribution of damage dose, represented by dpa, in the irradiation of 10.8 MeV tungsten ions on tungsten and corresponding number of collision cascades, induced by 10 keV PKAs, in MD simulations. The maximum damage dose is 0.02 dpa.}
  \label{Fig:dpa}
\end{figure}

Molecular dynamics (MD) simulations were performed to provide simulated targets for the RBS/c simulations. RBS/c is a depth-resolved technique. Hence, in order to reproduce experimental signals, the size of the simulated target along the probing direction, i.e., \hkl<111> direction, must be comparable to the thickness of damaged zone induced by the irradiation of 10.8 MeV tungsten ions. Due to computational limitations, the size of MD cells used for the generation of radiation damage is usually on the order of magnitude of 10 nm. However, as shown in Fig. \ref{Fig:dpa}, according to the SRIM calculations, the damaged region (blue dots) induced by the irradiation of 10.8 MeV tungsten ions on tungsten is beyond 1 $\mu$m (In the SRIM calculations, the "Quick calculation of damage" option was applied, and the threshold displacement energy of tungsten atom was set to 90 eV \cite{Banisalman2017}). In order to address this scale disparity resulted from the size limitation in radiation damage simulations and the extended damaged zone observed in the experiments, the generation of simulated targets is divided to two steps: generation of radiation damage in individual MD cells and construction of super-cells. 

\subsection{Generation of radiation damage}

The simulation of radiation damage in tungsten, which has a BCC lattice structure, was performed by using the PARCAS code \cite{Ghaly1999, Nordlund1995, Nordlund1998}. The MD cells were cuboid-shaped ($\sim$ 20 nm edge length). Periodic boundary conditions were applied to all three directions. The \hkl<111> direction of the tungsten lattice was aligned with the $z-$direction. Radiation damage was created by overlapping collision cascades, which was initiated by a primary knock-on atom (PKA) with an energy of 10 keV. We introduced collision cascades to two sets of MD cells to simulate radiation damage at 300 K and 800 K, respectively. Simulating radiation damage at a certain temperature refers to the procedure that the MD cells were thermally relaxed at the specified temperature before and after the introduction of collision cascades. Each set contained dozens of MD cells and each cell had different number of collision cascades, thus different level of radiation damage. The potential by Mason et al. \cite{Mason2023} was used for all steps of the MD simulations, including the generation of radiation damage in this step and the connection of MD cells in the next step. More detailed description of the methods for the generation of radiation damage, including the PKA simulation and relaxation simulation, can be found from Ref. \citenum{Granberg2021}.

\bigbreak

In order to quantitatively compare the radiation damage in MD simulations with that from SRIM calculations, we calculated the damage dose in the MD cells based on the Norgett-Robinson-Torrens (NRT) equation \cite{Norgett1975} as follows:
\begin{equation}
\text{dpa} =  \frac{0.8 E_{\text{$\nu$}}}{2 E_{\text{d}} N_{\text{tot}}} N_{\text{cas}} \label{Eq:nrt}
\end{equation}
where $N_{\text{cas}}$ represents the number of collision cascades, $N_{\text{tot}}$ represents the total number of atoms in one MD cell (492 480 atoms), $E_{\text{$\nu$}}$ gives the damage energy which represents the PKA energy not lost to electrons, and $E_{\text{d}}$ is the threshold displacement energy of tungsten atom (90 eV \cite{Banisalman2017}). For the PKA with an energy of 10 keV, $E_{\text{$\nu$}}$ equals to 7.8 keV. The damage dose in our MD cells, according to the NRT equation, ranges from 0 to 0.2 dpa. This range is large enough for comparing with the ion irradiation experiments, in which the maximum damage dose is only 0.02 dpa. 

\subsection{Construction of super-cells}

\begin{figure}[h!]
  \centering
  \includegraphics[width=0.9\linewidth]{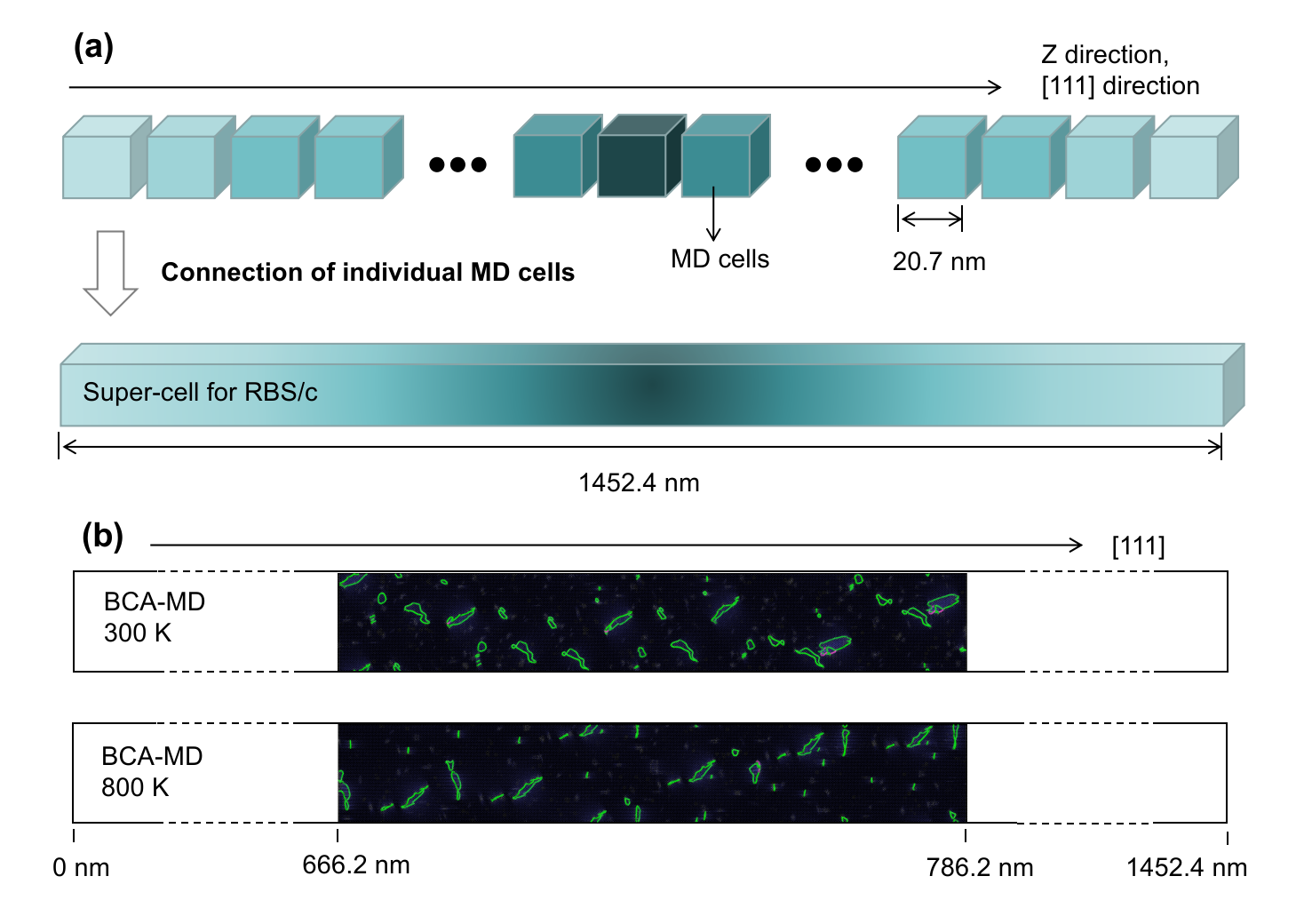}
  \caption{Construction of super-cells: (a) illustration for the assembly of individual MD cells along the \hkl<111> direction, and (b) visualization of dislocation structures in the super-cells (part) obtained via the BCA-MD approach at 300 K and 800 K.}
  \label{Fig:super-cell}
\end{figure}

After introducing radiation damage to the MD cells, these individual cells were assembled together in order to generate a simulated target suitable for the RBS/c simulations. More specifically, as shown in Fig. \ref{Fig:super-cell}(a), 70 MD cells were assembled together along the $z-$direction, i.e., the \hkl<111> direction, to form a super-cell whose side length along the $z-$direction (1.45 $\mu$m) can be comparable to the thickness of damaged zone in experiments.

\bigbreak

The arrangement of damaged MD cells along the $z-$direction follows two approaches:
\begin{enumerate}
\item In the first approach, the damage dose in each MD cell was firstly evaluated based on the number of collision cascades, using Eq.(\ref{Eq:nrt}). Then, the MD cells were arranged in such a way that the depth distribution of their damage dose follows that calculated by SRIM as shown in Fig. \ref{Fig:dpa}. Since SRIM calculations are based on the binary collision approximation (BCA), such approach is referred to as the BCA-MD approach. At each temperature, three sets of independent MD simulations were conducted to enhance statistical accuracy, and the resulting RBS/c spectra from these simulations were then averaged. Simulation results were compared with the experimental signals obtained from the damaged samples.

\item In the second approach, the MD cells were arranged such that the RBS/c signals generated from the super-cell could reproduce the experimental data. The exact configuration of the MD cells was determined through an iterative trial-and-error process: first, we fitted the experimental spectra in the high-energy region, then progressively adjusted the order of MD cells to match the spectra at lower energies, until the simulated spectra agreed with the experimental signals across the entire depth range. This approach is referred to as the Fit-MD approach. In this approach, one set of MD simulation was performed at each temperature. In addition to fitting the experimental signals obtained from the damaged samples, we also fitted the pristine spectrum. As the pristine sample may contain residual defects that can contribute to the yield in deep regions.

\end{enumerate} 
Fig. \ref{Fig:super-cell}(b) presents the visualization of dislocations within the depth range of 666.2 nm to 786.2 nm in the super-cells. The super-cells were constructed following the BCA-MD approach at 300 K and 800 K. The dislocations were identified by the dislocation analysis method implemented in the OVITO software \cite{Stukowski2012}. The observed dislocations are interstitial-type, consisting mainly of 1/2\hkl<111> loops (green) with a smaller fraction of \hkl<100> loops (red). The visualization region consists of 6 MD cells, and we can find that these MD cells were seamlessly connected.

\bigbreak

\begin{figure}[h!]
  \centering
  \includegraphics[width=0.9\linewidth]{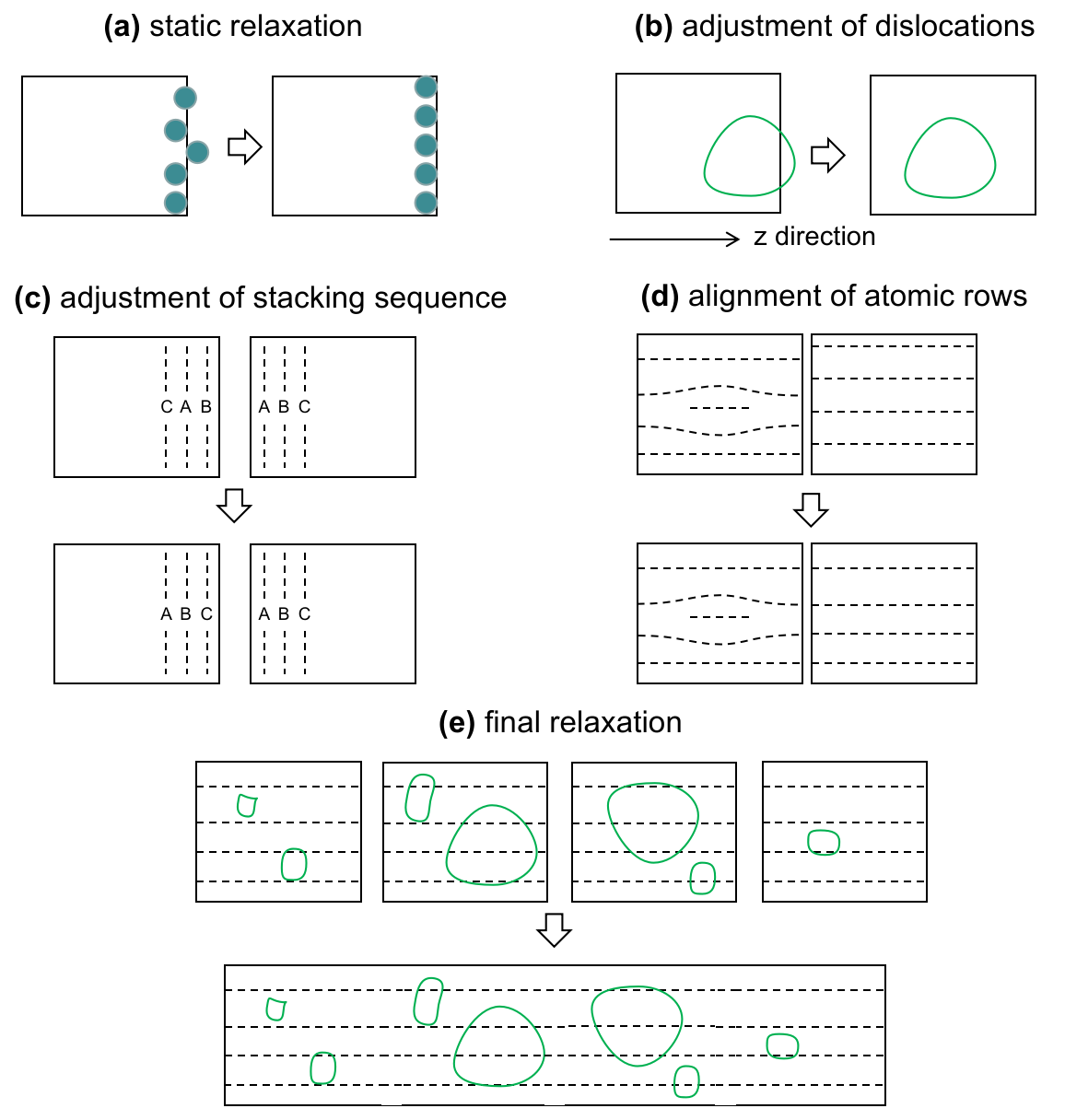}
  \caption{Illustration of MD cell connection steps}
  \label{Fig:cell_connect}
\end{figure}

The assembly of the MD cells along the depth direction is the first step in the super-cell construction. To achieve the super-cell configuration as shown in Fig. \ref{Fig:super-cell}(b), the MD cells must also be finely adjusted and finally relaxed to enable atomic interactions across cell boundaries, thereby establishing proper connectivity. This step is crucial for super-cell formation, as improper connection can either cause fracture of the whole structure or creation of high density defects localized at the MD cell interfaces. The general method for the connection of MD cells follows our previous work \cite{Markelj2024}. In this study, we further improved our method so that all the MD cells can be perfectly connected. The detailed methods are as follows:
\begin{itemize}
\item All adjustments to the MD cells are implemented based on the periodic boundary conditions: We firstly translate all atoms for a specified distance along a chosen direction, and then wrap atoms, that move outside the cell boundary, back to the opposite side of the cell. This preserves the overall properties of the cell, while allowing precise control of atomic configurations at the boundaries.

Since the MD cells are relaxed at a specified temperature after the introduction of collision cascades, the atoms exhibit thermal vibrations. As shown in Fig. \ref{Fig:cell_connect}(a), translating atoms along one direction may bring an atomic plane extremely close to the cell boundary. In such cases, thermal vibrations can cause some atoms to cross the boundary. Under periodic boundary conditions, these atoms appear on the opposite side of the cell. While this poses no issue when simulating a single MD cell, problems arise when connecting it to neighboring cells: the boundary-crossing atoms can no longer interact with their original atomic plane but instead with atoms in adjacent cells. As a result, these atoms effectively become interstitials, and their original positions turn into vacancies, thereby introducing artificial defects in the connection process. To prevent this problem, we performed static relaxation to all the MD cells prior to the connection. This ensures that atoms belonging to the same plane are not separated by cell boundaries.

\item Similarly, dislocations should not intersect cell boundaries, which, otherwise, could artificially generate new dislocations during the connection. As shown in Fig. \ref{Fig:cell_connect}(b), we translate atoms prior to the connection to address this issue.

\item When connecting MD cells, as shown in Fig. \ref{Fig:cell_connect}(c), we also carefully adjusted the cell boundaries to avoid stacking faults at the interfaces between adjacent cells.

\item When the MD cells are viewed along a crystallographic direction, such as the \hkl<111> direction, the atoms appear organized in strings or rows. We aligned these atomic rows across adjacent cells. During this process, it is important to note that some defective structures, such as dislocation loops, may also have their constituent atoms arranged in rows. As shown in Fig. \ref{Fig:cell_connect}(d), care must be taken to distinguish these defect-related atomic rows from those in the pristine lattice, ensuring that only the regular atomic rows are aligned. 

\item After the adjustments of the MD cells described above were completed, it is important to note that the system still consisted of 70 separate MD cells. To combine these cells into a single structure, i.e., a super-cell, we relaxed the entire system at 300 K and 0 Pa for 60 ps using the MD code LAMMPS \cite{Plimpton1995, Thompson2022}. The same potential \cite{Mason2023} as the one used for the generation of radiation damage was applied. This allowed atoms in adjacent cells to establish interatomic interactions, resulting in the formation of a unified super-cell, as shown in Fig. \ref{Fig:cell_connect}(e).
\end{itemize}
Eventually, after quenching the super-cell to 0 K in 10 ps, the resulting structure was used as the simulated target for the RBS/c simulations.

\section{Calculation of dislocation loop sizes}

\begin{figure}[h!]
  \centering
  \includegraphics[width=0.7\linewidth]{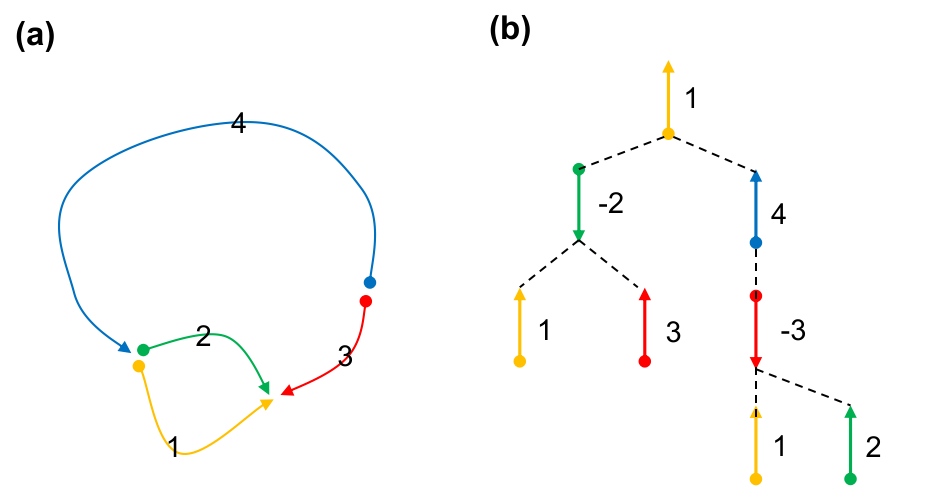}
  \caption{Identification of individual dislocation loops when multiple loops are in contact: (a) original dislocation configurations, (b) representation of original dislocation configurations as path-search problems.}
  \label{Fig:loop_iden}
\end{figure}

The dislocations in the MD cells within the damage dose range of 0 to 0.02 dpa were identified as dislocation loops using the OVITO software. For a single loop of length $L$, an equivalent loop diameter can be readily calculated as $D = L/\pi$. Nonetheless, the calculation of equivalent diameters is not always straight forward when loops are in contact. For example, as shown in Fig. \ref{Fig:loop_iden}(a), we can readily tell there are two loops in contact by visualization. However, it would be challenging to identify all the individual loops in the large amount of MD cells solely based on visualization, especially when multiple loops are in contact. In addition, the dislocation loops identified in the MD cells often appear in a piecewise form. For example, the small loop in Fig. \ref{Fig:loop_iden}(a) consists of two dislocation segments: segment No.1 and No.2, whereas the large loop consists of three dislocation segments: segment No.4, No.3 and No.2. This piecewise appearance of loop structures further complicates the problem.

\bigbreak

To facilitate the identification of individual loops, we re-framed the original task as finding a path that starts and ends at the same node, hence forming a closed loop, based on the depth-first search algorithm. Fig. \ref{Fig:loop_iden}(b) presents possible paths starting from the dislocation segment No.1, in which the start and end node of a segment are represented by an arrow and dot, respectively. The segment number is positive if the arrow points upward and negative otherwise. As shown in Fig. \ref{Fig:loop_iden}(b), the leftmost path going through the segments "1, -2, 1" can form a closed circle, hence enabling us to identify one loop. A further search along the second path "1, -2, 3" will be automatically terminated since we already found one loop in its parallel path. It seems that we can also identify a loop along the third path "1, 4, -3, 1". However, this loop will be abandoned since its perimeter is larger than that of the first path "1, -2, 1". As the case for the second path "1, -2, 3", the search along the last path "1, 4, -3, 2" will also be terminated. Eventually, the search starting from the segment No.1 identifies the small loop in Fig. \ref{Fig:loop_iden}(a). Whereas, a search starting from the segment No.3 can help us to found the large loop in Fig. \ref{Fig:loop_iden}(a). The equivalent loop diameter can be calculated as $\sum_{i} L_i / \pi$, where $L_i$ represents the length of dislocation segment $i$.

\bigbreak

\begin{figure}
     \centering
     \begin{subfigure}[b]{0.45\textwidth}
         \centering
         \includegraphics[width=\textwidth]{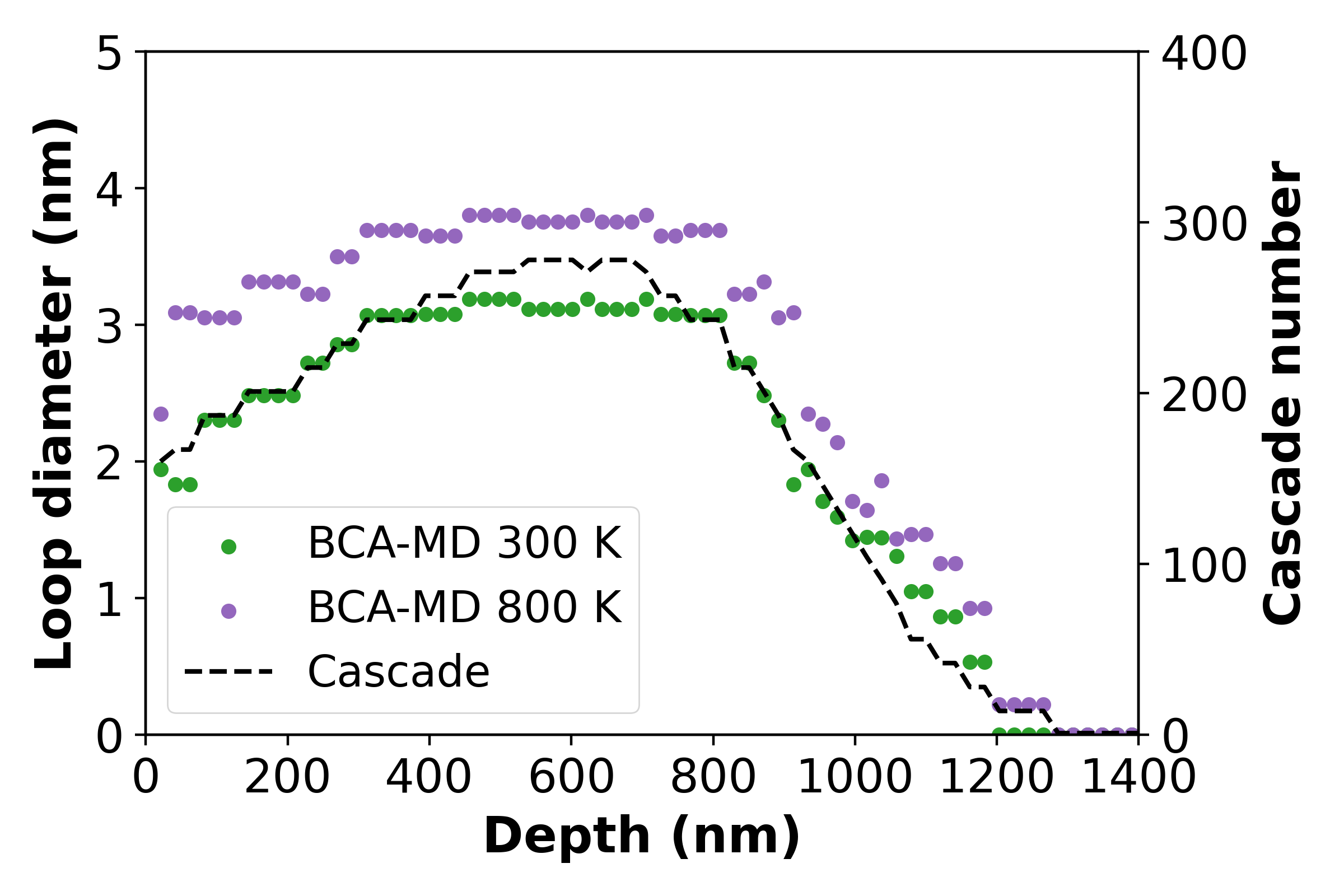}
         \caption{BCA-MD (weighted)}
     \end{subfigure}
     \begin{subfigure}[b]{0.45\textwidth}
         \centering
         \includegraphics[width=\textwidth]{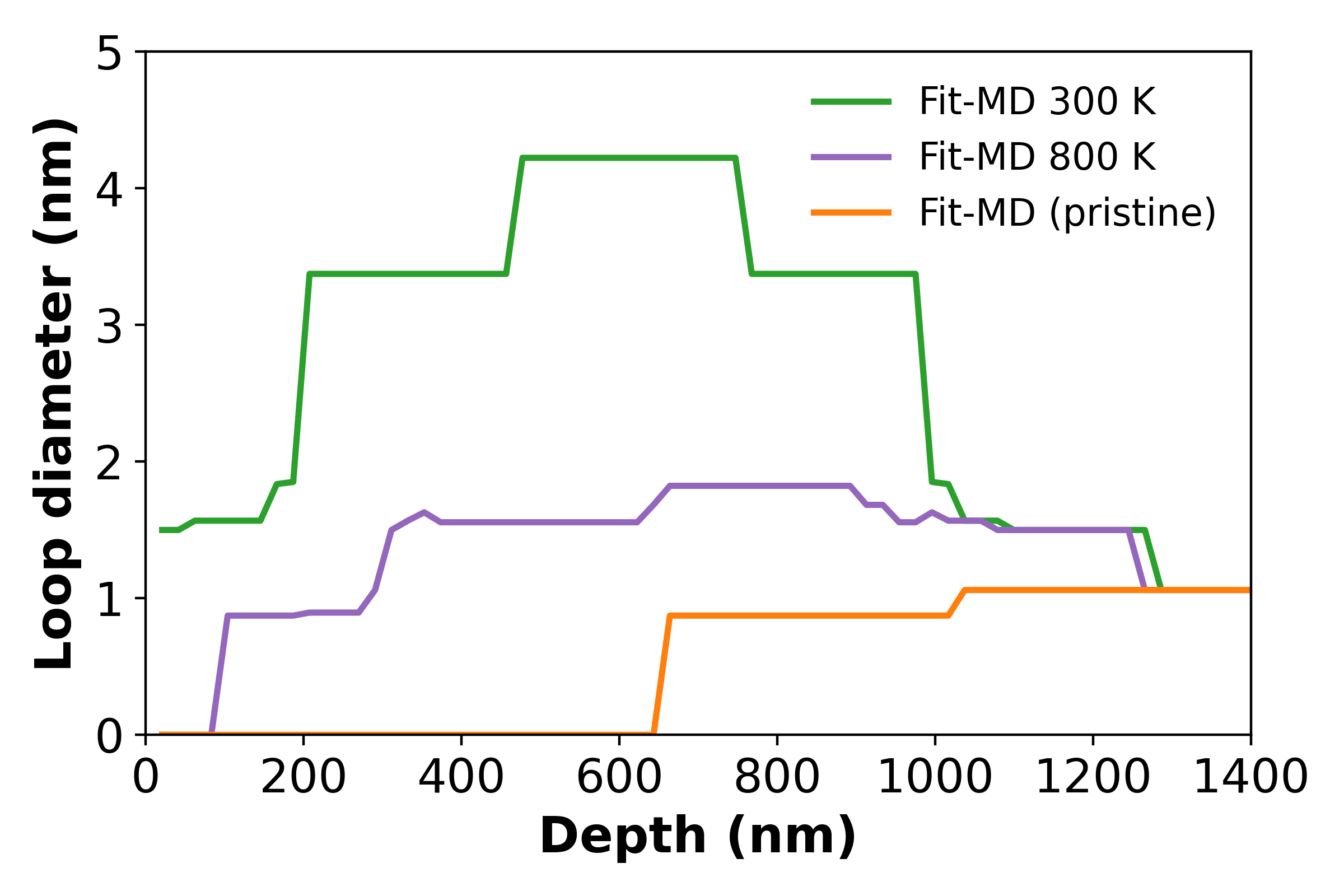}
         \caption{Fit-MD (weighted)}
     \end{subfigure}
        \caption{Depth distribution of loop diameters, calculated as the weighted average, in the simulated targets generated from the (a) BCA-MD approach and (b) Fit-MD approach.}
        \label{fig:loop_wei_ave}
\end{figure}

\begin{figure}
     \centering
     \begin{subfigure}[b]{0.45\textwidth}
         \centering
         \includegraphics[width=\textwidth]{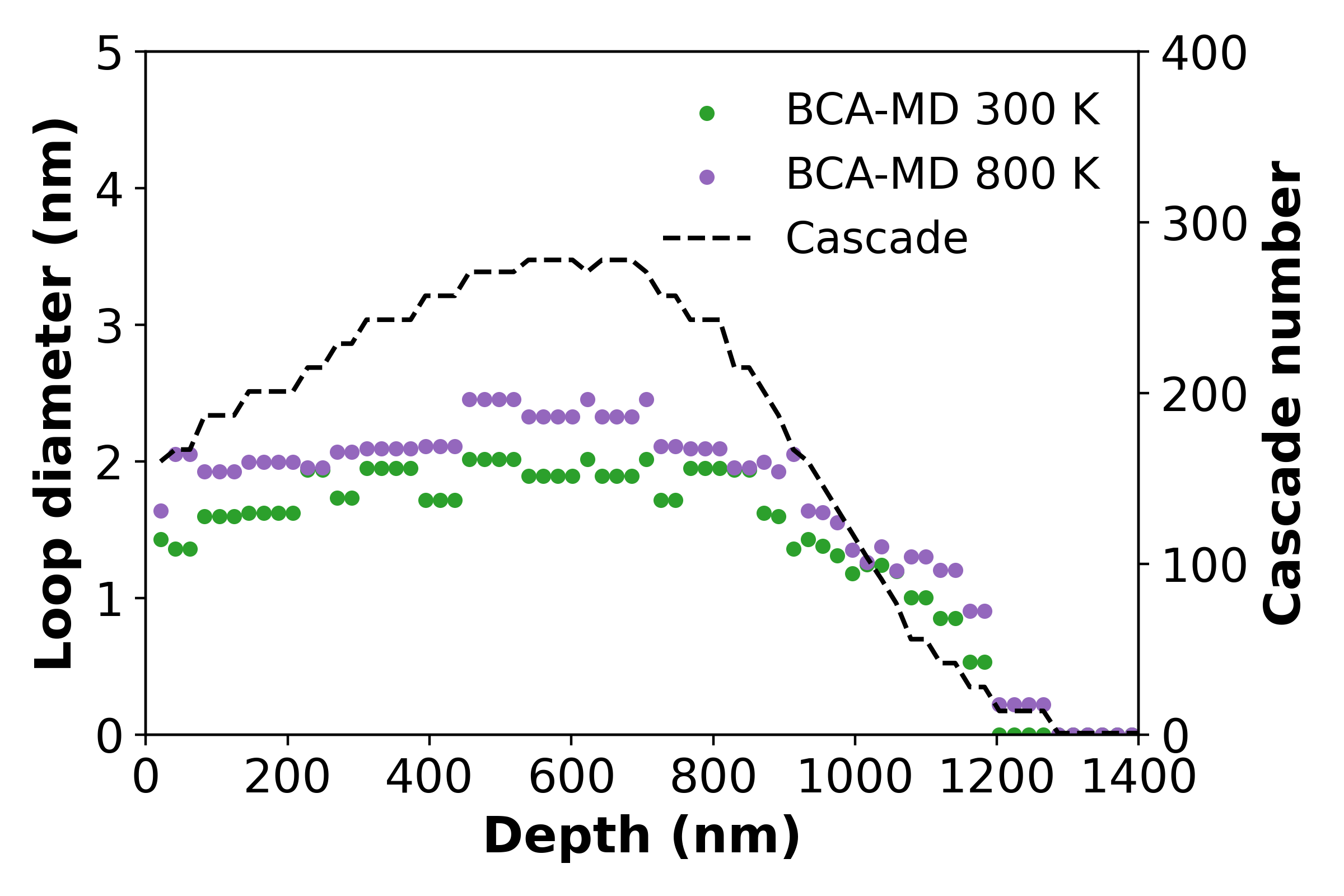}
         \caption{BCA-MD (arithmetic)}
     \end{subfigure}
     \begin{subfigure}[b]{0.45\textwidth}
         \centering
         \includegraphics[width=\textwidth]{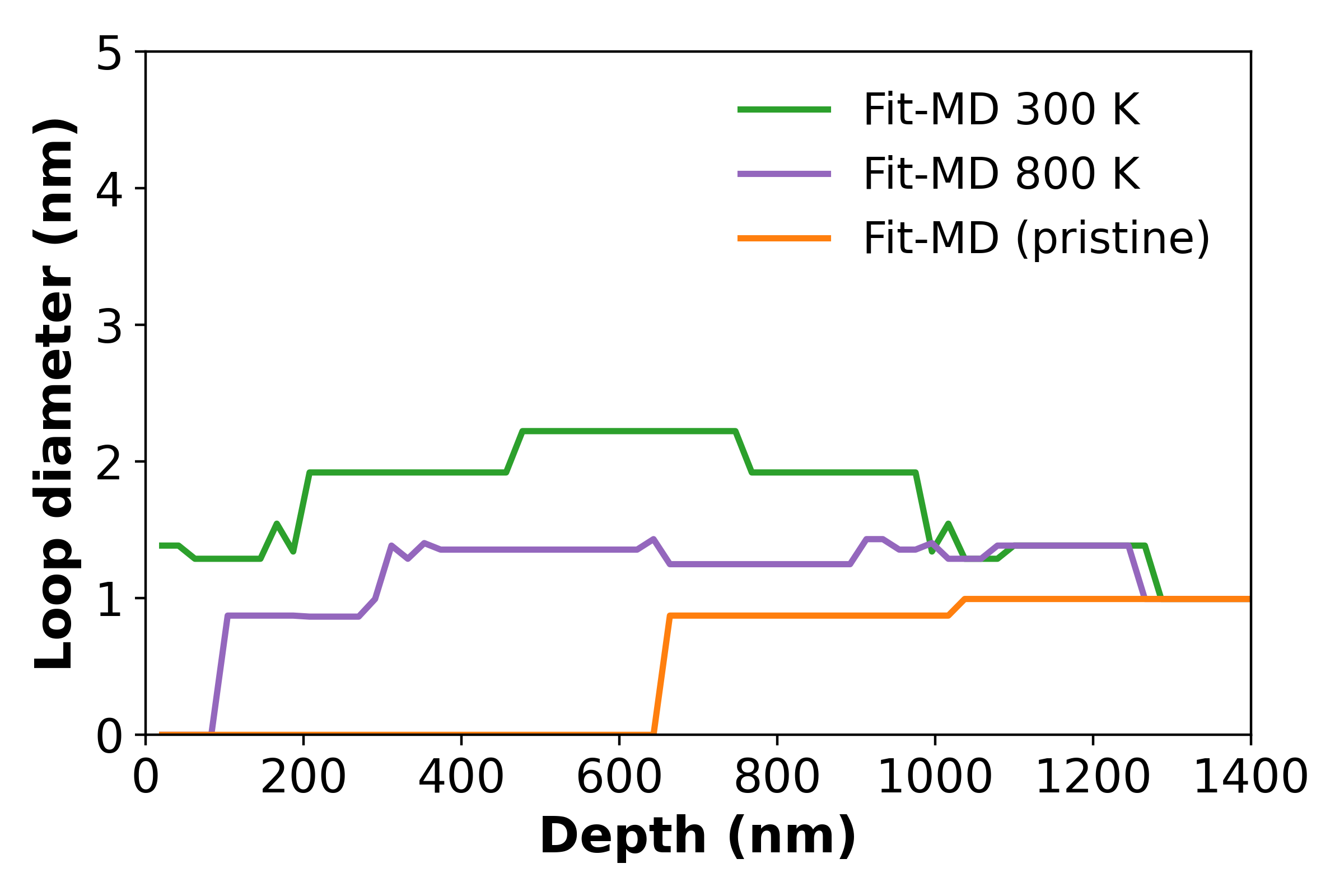}
         \caption{Fit-MD (arithmetic)}
     \end{subfigure}
        \caption{Depth distribution of loop diameters, calculated as the arithmetic average, in the simulated targets generated from the (a) BCA-MD approach and (b) Fit-MD approach.}
        \label{fig:loop_ave}
\end{figure}

Fig. \ref{fig:loop_wei_ave}(a) and (b) show the depth distribution of loop diameter in the simulated targets obtained from the BCA-MD and Fit-MD approaches, respectively. The loop sizes in these figures, $D_\text{ave}$, are average values weighted by the loop perimeter, $P$, which is calculated as follows:
\begin{equation}
D_\text{ave} = \frac{\sum_\text{n} P_\text{n} D_\text{n}}{\sum_\text{n} P_\text{n}} \label{Eq: dia_ave}
\end{equation}
where $D_\text{n}$ represents the equivalent diameter of loop $\text{n}$. We consider this weighted average, $D_\text{ave}$, to be preferable than an arithmetic average for two reasons: (i) larger loops could have a greater effect on RBS/c signals per unit loop length, (ii) small loops with diameters around 1 to 2 nm may fall below the detection threshold of transmission electron microscopy, so the weighted value may better align with experimental observations. 

\bigbreak

Fig. \ref{fig:loop_wei_ave}(a) shows that, for the BCA-MD approach, within the first \mbox{860 nm} region, the loop diameter at RT varies mainly from 2 to 3 nm, while the loop diameters at 800 K are approximately 1 nm larger than those grown at RT. In fact, this temperature-dependent behavior of the dislocation loops, seen as a decrease in density but an increase in size, agrees well with the observations of transmission electron microscopy (TEM) \cite{Ferroni2015}. In terms of the results obtained from the Fit-MD approach, Fig. \ref{fig:loop_wei_ave}(b) shows that the loops at RT are larger than those at 800 K. The maximal loop diameters are now 4.2 nm and 2 nm at RT and at 800 K, respectively. Despite this difference regarding the loop size between the two approaches, we need to emphasize as in the main text that this difference does not affect the overall shape of the dislocation density profile, which remains the most important parameter in this study.

\bigbreak

The arithmetic average results are also shown in Fig. \ref{fig:loop_ave} for completeness, in which the results for the simulated targets obtained from the BCA-MD and Fit-MD approaches are given in Figs. \ref{fig:loop_ave}(a) and (b), respectively. It can be observed that the arithmetic values are smaller than the weighted values given in Fig. \ref{fig:loop_wei_ave}.

\section{Calculation of surface effects}

\subsection{Calculation of surface stress}
In the calculations of surface stress on dislocation loops, we considered the interaction of a free surface with a circular dislocation loop. The loop lies on a plane parallel to the surface and its Burgers vector is normal to the surface. An equation based on the elasticity theory, first derived by Baštecká \cite{Bastecka1964} and further corrected by Fikar and Gröger \cite{Fikar2015}, was used to calculate the surface force on the loop. The surface force per unit length of the loop, $f_s$, as a function of the distance from the loop to the surface, $x$, can be expressed as:

\begin{equation}
f_\text{s} (x) = \frac{1}{8 \pi} \frac{b^2 \mu}{R (1 - \nu)} \left[ \frac{-3 d_\text{0}^4 + 7 d_\text{0}^2 + 2}{d_\text{0} (\sqrt{1 + d_\text{0}^2})^5} E_\text{I} \left( \frac{1}{1 + d_\text{0}^2} \right) + \frac{d_\text{0} (3 d_\text{0}^2 - 1)}{(\sqrt{1 + d_\text{0}^2})^5} K_\text{I} \left( \frac{1}{1 + d_\text{0}^2} \right) \right] \label{Eq:surf_stress}
\end{equation}
in which $R$ is the loop radius, $d_\text{0}$ equals to $x / R$, $b$ is the magnitude of the Burgers vector of the dislocation loop, $\mu$ is the shear modulus, $\nu$ is the Poisson ratio, $K_\text{I}$ and $E_\text{I}$ represents the complete elliptic integrals of the first and second kind, respectively. (Note that, in the original equation corrected by Fikar and Gröger \cite{Fikar2015}, $K_\text{I}$ and $E_\text{I}$ represents the complete elliptic integrals of the second and first kind, respectively. However, we found that, by setting $K_\text{I}$ and $E_\text{I}$ being the first and second kind, we can better reproduce the results in Bastecka \cite{Bastecka1964}.) The shear modulus, $\mu$, is calculated as follows \cite{Davis1998}:

\begin{equation}
\mu = \frac{E_\text{Y}}{2 (1 + \nu)}   \label{Eq:shear_modu}
\end{equation} 
where $E_\text{Y}$ is Young's modulus. The surface stress, $\tau_\text{s}$, is calculated by:

\begin{equation}
\tau_\text{s} = f_\text{s} / b
\end{equation}

\subsection{Temperature effects on surface stress}

\begin{figure}[h!]
  \centering
  \includegraphics[width=0.6\linewidth]{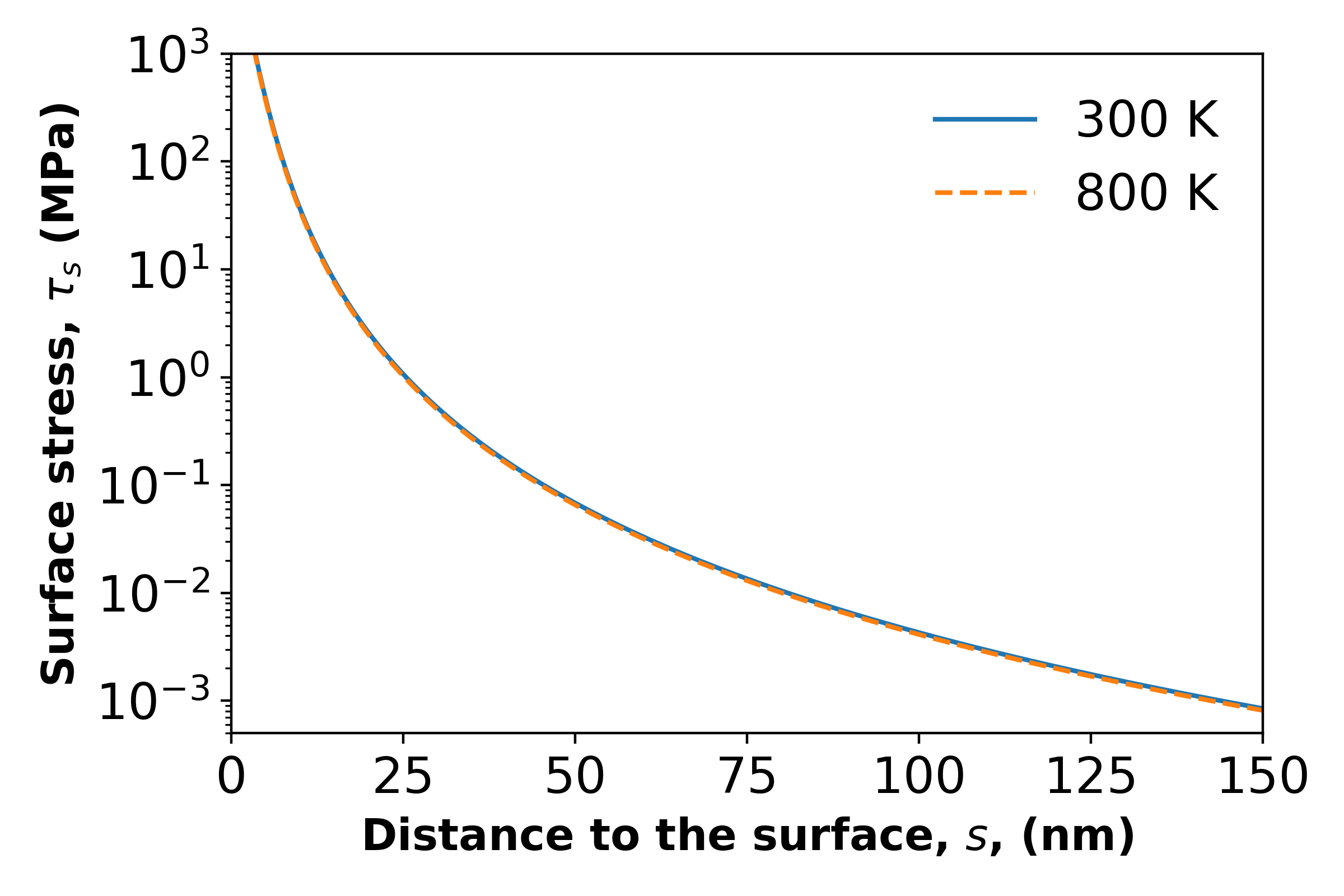}
  \caption{The surface stress on a 1/2\hkl<111> dislocation loop in tungsten at 300 K and 800 K.}
  \label{Fig:stress_temp}
\end{figure}

Since both $E_\text{Y}$ and $\nu$ can vary with temperature, the surface stress, $\tau_\text{s}$, can be affected by the temperature. Nonetheless, here we show that the effect of temperature (from 300 K to 800 K) is negligible for $\tau_\text{s}$ in tungsten. For tungsten, the variation of $\nu$ with temperature, $T$ (in Kelvin), is obtained from this equation \cite{Skoro2011}:

\begin{equation}
\nu = 0.279 + 1.0893 \times 10^{-5} \cdot (T - 273.15) \label{Eq:poisson_tem}
\end{equation} 
The experimental measurements of static $E_\text{Y}$ of tungsten at 300 K and at a temperature close to 800 K (810 K) are 395 GPa and 378 Gpa, respectively \cite{Harrigill1972}. The lattice parameter of tungsten, $a_0$, is 3.165 Å. Using Eq.(\ref{Eq:surf_stress}-\ref{Eq:poisson_tem}) and the experimental values of $E_\text{Y}$, $\tau_\text{s}$ on a circular dislocation loop at 300 K and 800 K can be calculated. Fig. \ref{Fig:stress_temp} shows the surface stress on a circular 1/2\hkl<111> dislocation loop in tungsten at 300 K and 800 K. The radius of the loop is 5 nm as in the main text. With a larger distance from the loop to the surface, $\tau_\text{s}$ decreases rapidly. Nonetheless, the difference between the stress at 300 K and 800 K is rather small. Hence, the effect of temperature on the stress is negligible. The surface stress given in the main text was calculated at 300 K. 

\subsection{Calculation for the drift movement of dislocation loops}

Under an external force, $F$, dislocation loops can experience a drift motion. Such force can produce an applied stress, $\tau$, on a dislocation loop as follows:

\begin{equation}
\tau = \frac{F}{2 \pi R b} \label{Eq:applied_stress}
\end{equation}
The drift velocity, $v_D$, is calculated by \cite{Hirth1982}:

\begin{equation}
v_\text{D} = F \frac{D_\text{eff}}{K T} \label{Eq:drift_v}
\end{equation}
where $D_\text{eff}$ is the diffusion coefficient of the loop, $K$ is Boltzmann constant and $D_\text{eff}/KT$ represents the mobility, $u$. The $D_\text{eff}$ is calculated by:

\begin{equation}
D_\text{eff} = \frac{1}{2} b^2 \frac{\nu_\text{0}}{\sqrt{N}} \text{exp} \left( \frac{-(E_\text{m} - E_\text{s})}{KT} \right) \label{Eq:diffusion_coef}
\end{equation}
where $\nu_\text{0}$ is the attempt frequency of jump over a distance of $b$ (in this work, $\nu_\text{0} = 10^{13}$ s$^{-1}$, as in Ref. \citenum{Renterghem2022}), $N$ is the number of interstitial atoms forming the dislocation loop, $E_\text{m}$ is the effective migration energy and $E_\text{s}$ represents the interaction energy induced by the applied stress. The constant 1/2 in Eq.(\ref{Eq:diffusion_coef}) comes from the fact that we are assuming a one dimensional motion of the loop. For a 1/2\hkl<111> dislocation loop, $N$ can be obtained from the following equation as proposed in Ref. \citenum{Bakaev2021}:

\begin{equation}
N = \sqrt{3} \pi \frac{R^2}{a_0^2} \label{Eq:sia_num}
\end{equation}
$E_\text{s}$ is calculated according to Ref. \citenum{Jager1975}:

\begin{equation}
E_\text{s} = \frac{1}{2} \pi R^2 b \tau \label{Eq:surf_ener}
\end{equation} 
The drift velocity of loops increases when it approaches the open surface, since both the surface force and then diffusion coefficient get larger (note that the surface energy affects the diffusion coefficient, and this is why the diffusion coefficient changes). The time, $t$, required for a loop to cross a certain distance, $x_\text{s}$, before reaching the open surface can be calculated as follows:
\begin{equation}
t = \int_0^{x_s} \frac{dx}{v(x)} = \int_0^{x_s} \frac{K T}{2 \pi R f_s(x) D_\text{eff}(x)}dx \label{Eq:int}
\end{equation} 
This integration can be solved by numerical methods, such as the Riemann sums.

\begin{figure}[h!]
  \centering
  \includegraphics[width=0.6\linewidth]{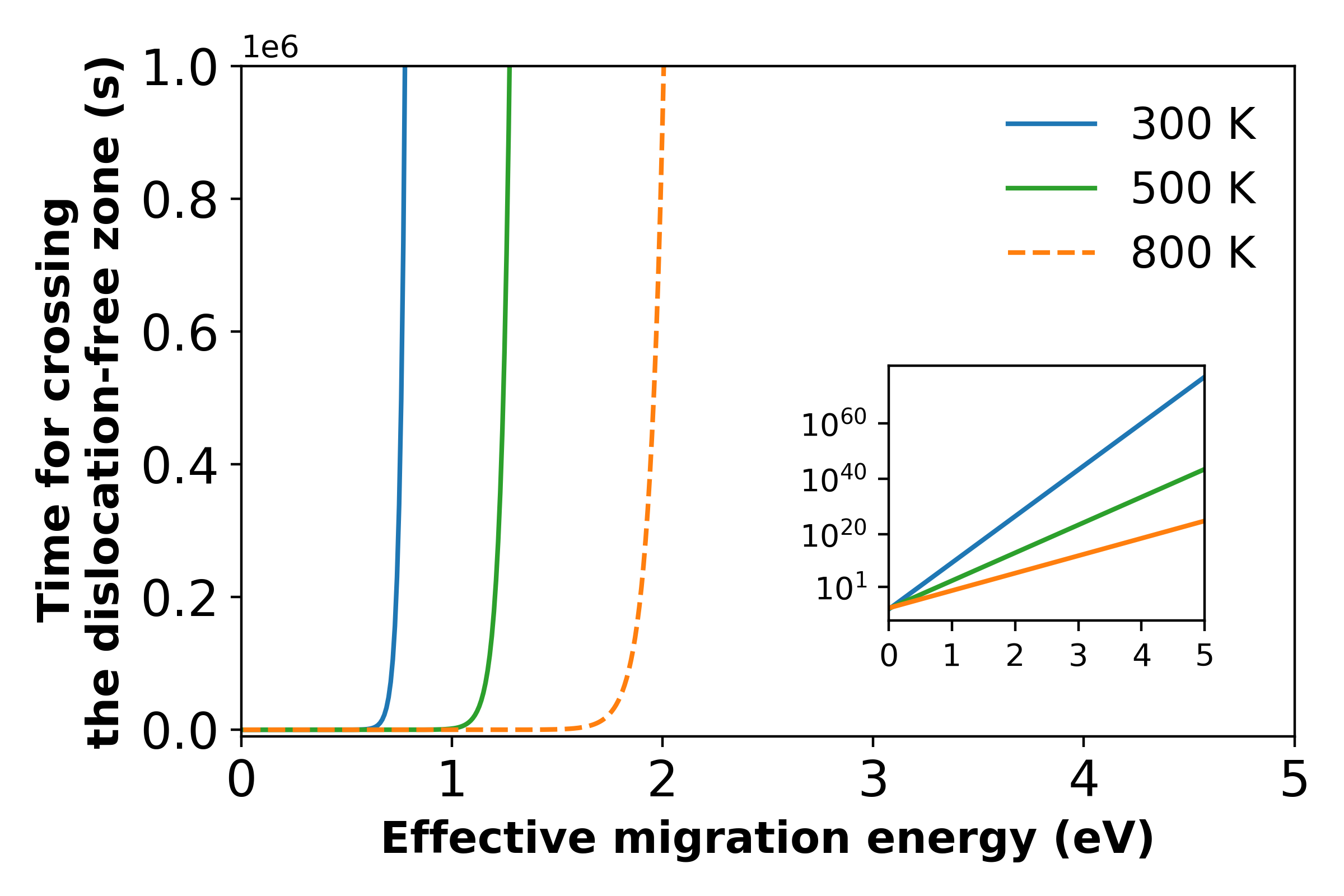}
  \caption{Time required for a dislocation loop to cross the dislocation-free zone (83 nm) as a function of the effective migration energy. The inset shows the corresponding time on a logarithmic scale.}
  \label{Fig:drift_v}
\end{figure}

\bigbreak

Figure. \ref{Fig:drift_v} shows the time required for a 1/2\hkl<111> dislocation loop to cross the dislocation-free zone (83 nm) as a function of the effective migration energy. The loop diameter is 5 nm as in the main text. We can observe that the drifting time increases with a higher migration energy. If the effective migration energy is higher than a certain value, we can consider that the loop at the boundary of dislocation-free zone (83 nm) will never reach the surface. For example, at 300 K, if the effective migration energy is larger than 1 eV, the calculated time would be extremely large.   

\bibliographystyle{naturemag}
\bibliography{./References/MyLibrary.bib,./References/Non_downloaded.bib}
